\documentclass[twoside,10pt]{article}
\usepackage{epsfig}

\newcommand{\be}{\begin{equation}}
\newcommand{\ee}{\end{equation}}
\newcommand{\bea}{\begin{eqnarray}}
\newcommand{\eea}{\end{eqnarray}}

\begin{document}

\title{Direct dark matter detection: The diurnal variation in directional
      experiments }

\author{J.D. Vergados$^{1}$ and Ch.C. Moustakidis$^{2}$ \\
$^{1}$Theoretical Physics Division, University
of Ioannina, Ioannina, Gr 451 10, Greece \\
$^{2}$Department of Theoretical Physics, Aristotle University of
Thessaloniki, \\ 54124 Thessaloniki, Greece
}

\maketitle

\begin{abstract}
We present some theoretical results relevant to the direct dark matter
detection experiments, paying particular attention to directional
experiments,  i.e. experiments in which, not only the energy but
the direction of the recoiling nucleus is observed.
In directional experiments the detection rate depends on the angle
between the line observation and the sun's direction of motion. Since, however, the
direction of observation is fixed with respect the earth, while
the Earth is rotating around its axis, in a directional experiment
the angle between the direction of observation and the Sun's
direction of motion will change during the day. So the observed signal
in such experiments will exhibit a very interesting and
characteristic periodic diurnal variation.
\\
\\

PACS number(s): 95.35.+d, 12.60.Jv

Keywords: Dark matter, Supersymmetric models
\end{abstract}

\section{Introduction}


The combined MAXIMA-1 \cite{MAXIMA-1}, BOOMERANG \cite{BOOMERANG},
DASI \cite{DASI} and COBE/DMR Cosmic Microwave Background (CMB)
observations \cite{COBE} imply that the Universe is flat
\cite{flat01}
and that most of the matter in the Universe is Dark
\cite{SPERGEL},  i.e. exotic. These results have been confirmed
and improved by the recent WMAP data \cite{WMAP06}. Combining the
the data of these quite precise experiments one finds:
$$\Omega_b=0.0456 \pm 0.0015, \quad \Omega _{CDM}=0.228 \pm 0.013 , \quad \Omega_{\Lambda}= 0.726 \pm 0.015.$$
Since any "invisible" non exotic component cannot possibly exceed
$40\%$ of the above $ \Omega _{CDM}$ ~\cite {Benne}, exotic (non
baryonic) matter is required and there is room for cold dark
matter candidates or WIMPs (Weakly Interacting Massive Particles).

Since the WIMP's are  expected to be very massive
and extremely non relativistic,  they are not likely to excite the nucleus. So they can be
directly detected mainly via the recoiling of a nucleus (A,Z) in
elastic scattering.
In the standard nuclear recoil experiments, first proposed more
than 30 years ago \cite{GOODWIT}, one has to face the problem that
the reaction of interest does not have a characteristic feature to
distinguish it from the background. So for the expected low
counting rates the background is a formidable problem. Some
special features of the WIMP-nuclear interaction can be exploited
to reduce the background problems. Such are: i) the modulation
effect: this yields a periodic signal due to the motion of the
earth around the sun. ii) backward-forward asymmetry expected in directional
experiments, i.e. experiments in which the direction of the
recoiling nucleus is also observed. iii)
transitions to very low excited states, if present in the target: in this case one need not measure
nuclear recoils, but the de-excitation $\gamma$ rays. (iv)
detection of electrons produced during the WIMP-nucleus collision or subsequent X-rays.

In our previous work \cite{JDV03,JDVSPIN04,VF07} we have found that the observed directional rate is characterized by a strong dependence
on the angle between the direction of nuclear recoil and the direction of the sun's motion, with respect to the center of the galaxy.
The apparatus will, of course, be oriented in a direction defined
in the local frame, i.e.  the line of observation will point to a
point in the sky  specified, in the equatorial system, by right
ascension $\tilde{\alpha}$ and inclination $\tilde{\delta}$. Thus
due to the rotation of the earth in a  directional experiment the
angle between the direction of observation and the Sun's direction
of motion will change during the day. So, since the event rates
sensitively depend on this angle, as we have recently
suggested~\cite{CYGNUS09} one may be able to observe a diurnal
variation.
In the present paper we will explore this novel feature of directional experiments,
 i.e. the variation of the data with a period of 24 hours.
This variation is independent of the Earth's rotational velocity. Its amplitude depends, in addition
to some of the usual parameters specifying   the standard WIMP event rate, on the inclination of observation, i.e.
the angle between the direction of observation and the axis of the Earth's rotation.
Those features cannot be masked by any known background.


\section{Standard (non Directional) Rates}
The event rate for a given WIMP velocity $\upsilon$ is given  by \cite{Vergados-010}
\begin{equation}
\frac{{\rm d} R}{ {\rm d}u}=\frac{\rho_{\chi}}{m_{\chi}} \frac{m_t}{A m_p}\sqrt{<\upsilon^2>} A^2  \sigma_n \left ( \frac{\mu_r}{\mu_p} \right )^2 \frac{{\rm d} t(u,\upsilon)}{{\rm d} u},
\label{Eq:dRdu}
\end{equation}
where ${m_t}/{A m_p}$ is the number of nuclei in a target of mass $m_t$, $A$ is the mass number of the target, $u$ is the energy transfer in some convenient units,
$u={Q}/{Q_0}~,~Q_0=40 A^{-4/3} \mbox { MeV }$, and
\begin{equation}
\frac{{\rm d} t(u,\upsilon)}{{\rm d} u}=\frac{\upsilon}{\sqrt{<\upsilon^2>}} F^2(u) \frac{1}{2 (\mu_r b \upsilon)^2},
\label{Eq:dtdu}
\end{equation}
with $F(u)$ the nuclear form factor. We must now fold ${\rm d} t(u,\upsilon)/{\rm d}u$ with the velocity distribution $f(y,\xi)$ in the local frame. Here we use $y=\upsilon/\upsilon_0$, $\upsilon_0$ is the sun's velocity with respect to the center of the galaxy, and $\xi$ is the cosine of the the WIMP velocity with the polar axis (chosen to be the sun's direction of motion).
We  will  assume in this work that the velocity distribution is Maxwell-Boltzmann in the galactic frame, namely
\begin{equation}
f({\bf y}')=\frac{1}{ \pi \sqrt{\pi}} e^{-({\bf y}')^2}, \qquad {\bf y}'=\frac{\bf{\upsilon}'}{\upsilon_0},
\label{Eq:M-B}
\end{equation}
where $\bf{\upsilon}'$ is the WIMP velocity in the galactic frame.
 Obviously to obtain  the distribution in the local frame one must take into account the motion of the Sun and Earth:
\begin{equation}
{\bf y}'={\bf y}+{\hat z}+\delta \left (\sin{\alpha}{\hat x}-\cos{\alpha}\cos{\gamma}{\hat y}+\cos{\alpha}\sin{\gamma} {\hat z}\right ),
\end{equation}
where $\delta=\upsilon_1/\upsilon_0=0.136$ with $\upsilon_1$ the velocity of the earth around the sun, $\alpha$ the phase of the earth ($\alpha$=0 on June 3rd) and $\gamma\approx\pi/6$.

In the case where $\delta=0$,  we get for the differential (with respect to the energy transfer) rate
\begin{equation}
\frac{{\rm d} R}{ {\rm d} u}=\frac{\rho_{\chi}}{m_{\chi}}\frac{m_t}{A m_p} \sigma_n \left ( \frac{\mu_r}{\mu_p} \right )^2 \sqrt{< \upsilon^2 > }A^2\frac{d t}{du},\qquad \frac{{\rm d} t}{{\rm d} u}=\sqrt{\frac{2}{3}} a^2 F^2(u) \Psi_0(a \sqrt{u}),
\label{DR-1}
\end{equation}
\begin{equation}
 \Psi_0(a \sqrt{u})=\int_{a\sqrt{u}} ^{y_{esc}} y {\rm d}y~ 2 \pi \int_{-1}^{1} {\rm d} \xi f(y,\xi),
\end{equation}
where
\begin{equation}
a=(\sqrt{2} \mu_r b \upsilon_0)^{-1}.
\end{equation}
We note that the function $\Psi_0(x)$,  a decreasing function of the energy transfer, depends only on the velocity distribution and is independent of nuclear physics.
An additional suppression as the energy transfer increases comes, of course,
from the nuclear form factor $ F(u)$.

 Note that the nuclear dependence of the differential rate comes mainly from the form factor and, to some extent,  via the parameter $a$.
Integrating the differential event rate from $u_{min}=E_{th}/Q_0$, which depends on the energy threshold, to $ u_{max}=y^2_{esc}/a^2$ (where $y_{esc}=2.84$ since $\upsilon_{esc}=2.84 \upsilon_0$) we obtain the total rate
\begin{equation}
R=\frac{\rho_{\chi}}{m_{\chi}}\frac{m_t}{A m_p} \sigma_n  \left ( \frac{\mu_r}{\mu_p} \right )^2 \sqrt{< \upsilon^2 > }A^2t,
\qquad t=\int_{u_{min}} ^{u_{max}} {\rm d} u\frac{{\rm d} t}{{\rm d} u}.
\end{equation}

%


In the case where $\delta \neq 0$ we proceed as above and finally get for the differential rate an expression similar to (\ref{DR-1}) where now one has to replace:
\begin{equation}
\Psi_0(x)\rightarrow\Psi_0(x)+\Psi_1(x,\gamma,\delta) \cos{\alpha},
\end{equation}
where $\Psi_1(x,\gamma,\delta)$ is obtained as the linear term in an expansion of the velocity distribution in powers od $\delta$. Thus
one has to replace:

\begin{equation}
\frac{{\rm d} t}{{\rm d} u}\rightarrow\frac{{\rm d} r}{{\rm d} u}=\frac{{\rm d} t}{{\rm d} u}+\frac{{\rm d} \tilde{h}}{{\rm d} u} \cos{\alpha},
\end{equation}
\begin{equation}
t\rightarrow t(1+h\cos{\alpha}), \quad h=\frac{1}{t}\int_{u_{min}}^{u_{max}} {\rm d}u \frac{{\rm d} \tilde{h}}{{\rm d} u}.
\end{equation}
Thus we finally get
\begin{equation}
R=\frac{\rho_{\chi}}{m_{\chi}}\frac{m_t}{A m_p} \sigma_n
\left ( \frac{\mu_r}{\mu_p} \right )^2 \sqrt{<\upsilon^2>}  A^2~t\left (1+h \cos{\alpha} \right ).
\label{Eq:totalR}
\end{equation}


 \begin{figure}
 \begin{center}
 {
\rotatebox{90}{\hspace{-0.0cm} {$h \longrightarrow$}}
\includegraphics[scale=0.5]{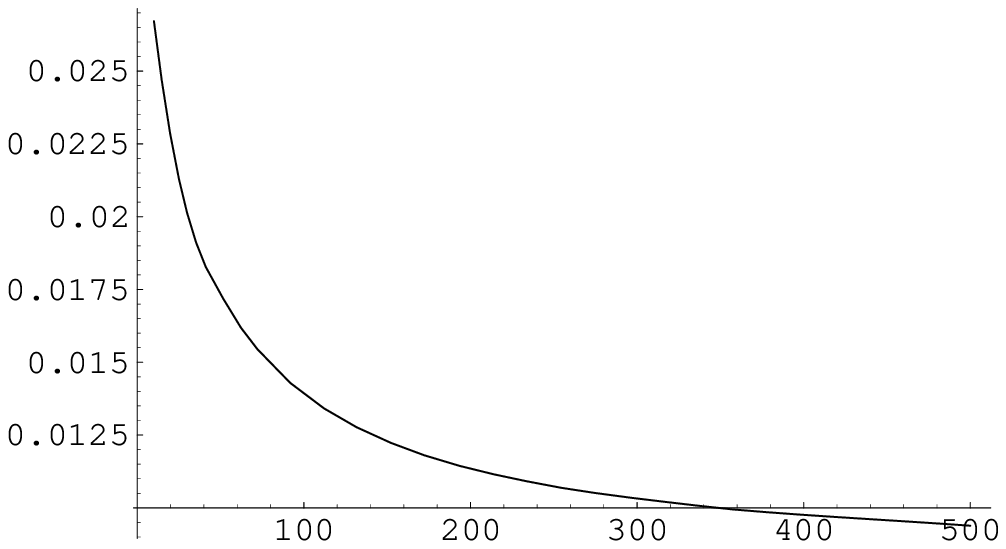}
}
 {
\rotatebox{90}{\hspace{-0.0cm} {$h \longrightarrow$}}
\includegraphics[scale=0.5]{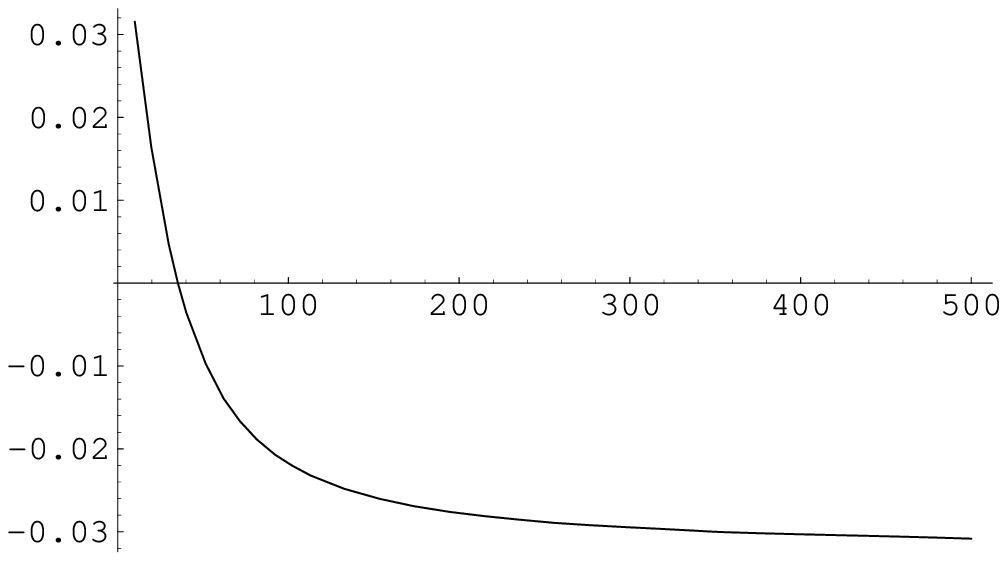}
}
\\
{\hspace{1.0cm} $m_{\chi}\longrightarrow $ GeV}
\caption{The quantity $h$, entering in the coherent mode discussed in this work, is shown as a function of the WIMP mass in GeV for  $Q_{min}=0$ for a light system, $^{32}$S (left) and  $^{127}$I or $^{131}$Xe (right).}
 \label{fig:totalh}
\end{center}
  \end{figure}

\begin{figure}
\begin{center}
 {
\rotatebox{90}{\hspace{-0.0cm} {$R \longrightarrow $ events/kg/y}}
\includegraphics[scale=0.5]{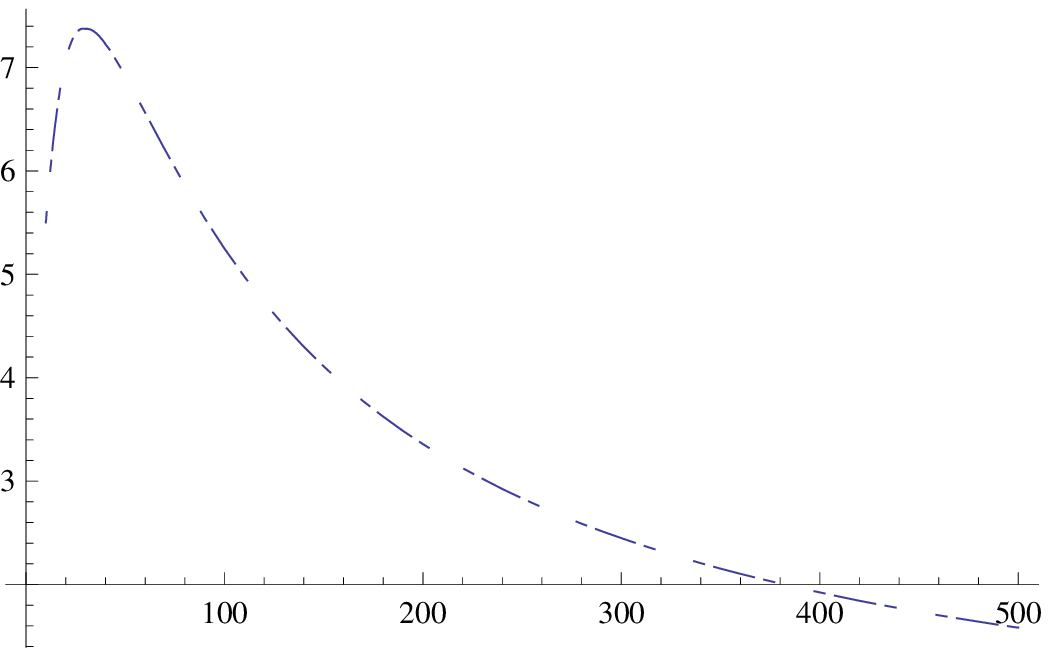}
}
 {
\rotatebox{90}{\hspace{-0.0cm} {$R \longrightarrow $ events/kg/y}}
\includegraphics[scale=0.5]{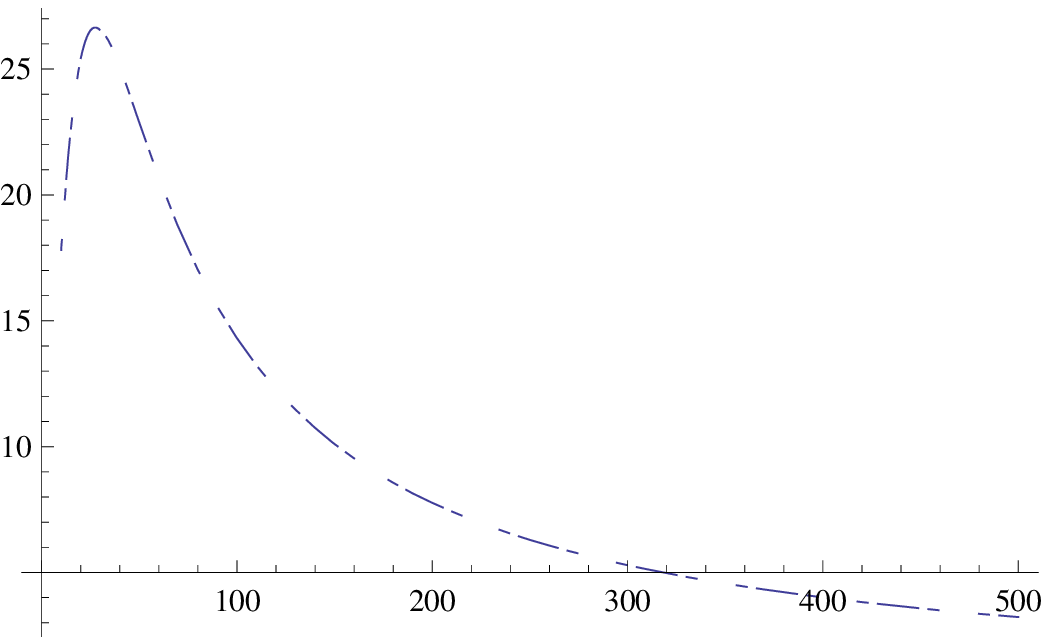}
}
\\
{\hspace{1.0cm} $m_{\chi}\longrightarrow $ GeV}
\caption{The coherent total event rate, as a function of the WIMP mass in GeV, is exhibited assuming a nucleon
 cross section $\sigma_n=10^{-43}$ cm$^2$=$10^{-7}$pb
  for $Q_{min}=0$
   for a light system, $^{32}$S (left) and a heavy system,
     $^{131}$Xe (right).}
 \label{fig:totalR}
\end{center}
\end{figure}


\section{Directional Rates}
In this instance the experiments will attempt to measure not only the energy, but the direction of the recoiling nucleus as well  \cite{SPERGEL-88,JDVSPIN04,VF07,Vergados-010}.
Let us indicate the direction of observation by:
\begin{equation}
{\hat e}=(e_x,e_y,e_z)=(\sin{\Theta}\cos{\Phi},\sin{\Theta}\sin{\Phi},\cos{\Theta}).
\end{equation}
Now Eq. (\ref{Eq:dtdu}) must be motified since  no integration over the azimuthal angle specifying the direction of the outgoing nucleus is performed. Thus Eqs.~(\ref{Eq:dRdu}) and (\ref{Eq:dtdu} ) now become
\begin{equation}
\left (\frac{{\rm d} R}{ {\rm d} u}\right )_{dir}=\frac{\rho_{\chi}}{m_{\chi}} \frac{m_t}{A m_p}\sqrt{<\upsilon^2>} A^2  \sigma_n \left ( \frac{\mu_r}{\mu_p} \right )^2 \frac{1}{2 \pi}\left (\frac{{\rm d} t(u,\upsilon)}{{\rm d} u} \right )_{dir},
\label{Eq:dirdRdu}
\end{equation}
\begin{equation}
\left (\frac{{\rm d} t(u,\upsilon)}{{\rm d} u}\right )_{dir}=\frac{\upsilon}{\sqrt{<\upsilon^2>}} F^2(u) \frac{1}{2 (\mu_r b \upsilon)^2}\delta\left ({\hat \upsilon}.{\hat e}-\frac{a \sqrt{u}}{\upsilon/\upsilon_0}\right ).
\label{Eq:dirdtdu}
\end{equation}
The factor $1/( 2 \pi) $ enters since we are going to make use of the same nuclear cross section as in the non directional case.


The total event rate is obtained after  folding with the velocity distribution and integrating the above expression over the energy transfer. It can be cast in the form:
\begin{eqnarray}
R_{dir}&=&\frac{\rho_{\chi}}{m_{\chi}}\frac{m_t}{A m_p} \sigma_n \left ( \frac{\mu_r}{\mu_p} \right )^2 \sqrt{< \upsilon^2 > }  A^2\frac{1}{2 \pi} t_{dir}(\Theta,\Phi)\nonumber\\
&&\left ( \frac{}{} 1+h_c(\Theta,\Phi) \cos{\alpha}+h_s(\Theta,\Phi) \sin {\alpha}\right ).
\end{eqnarray}
We found it convenient to use a new  relative parameter  $\kappa$\footnote{This is done to avoid or minimize, e.g., the dependence on the  parameters  of the particle model and, in particular, the unknown WIMP-nucleon cross section. Furthermore the directional experiments can also obtain the rate in all directions. Thus $\kappa$ maybe of experimental interest.} as well as the parameters $h_m$ and $\alpha_0$ given by:
\begin{equation}
\kappa(\Theta,\Phi)=\frac{t_{dir}}{t},\qquad   h_m(\Theta,\Phi)\cos{(\alpha+\alpha_0)}=h_c(\Theta,\Phi) \cos{\alpha}+h_s(\Theta,\Phi) \sin {\alpha}.
\end{equation}

These parameters depend, of course, on the reduced mass $\mu_r$, the WIMP velocity distribution and, to some extent, on the nuclear physics via the nuclear form factor.
The parameter $\kappa$ gives the retardation factor of the directional rate, over and above the factor of $1/(2 \pi)$, compared to the standard rate.  Since, however, the unmodulated amplitude is independent of $\Phi$ one can integrate over $\Phi$ so that the suppression of $1/(2 \pi)$ drops out for this term.
Because of the existence of both $\cos{\alpha}$ and $\sin{\alpha}$ terms the time dependence will be of the form $\cos{(\alpha+\alpha_0)}$, with the phase $\alpha_0 $ being direction dependent. Thus the time of the maximum and minimum will depend on the direction. In other words the seasonal dependence will depend on the direction of observation. So it cannot be masked by irrelevant seasonal effects.

Before concluding this section we like to consider  the case connected with the partly directional experiments, i.e. experiments which can determine the line along which the nucleus is recoiling, but not the sense of direction on it. The results in this case can be obtained  by summing up the events in both directions. i.e. those specified by $(\Theta,\Phi)$ as well as  $(\pi-\Theta,\Phi+\pi)$. A given line of observation is now specified by $\Theta$, $\Phi$ in the range:
\[  0 \leq \Theta \leq \pi/2, \qquad   0 \leq \Phi \leq \pi. \]

The parameter $h$ entering the non directional case is shown in Fig.~\ref{fig:totalh}.  Using  the above parameters $t$ and employing  Eq. (\ref{Eq:totalR}) we obtain the time independent total
 event rate. The obtained results are shown in Fig. \ref{fig:totalR}. The total modulation, since it is
 defined relative to the time independent part, is still given by Fig. \ref{fig:totalh}.

\begin{figure}
\begin{center}
\rotatebox{90}{\hspace{2.0cm} {$\Theta$ (radians)}}
\includegraphics[scale=0.6]{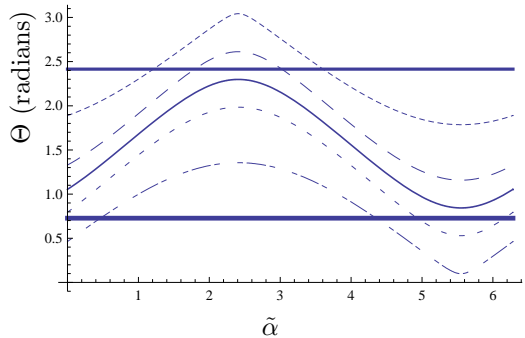}\\
\hspace{-0.0cm} {$\tilde{\alpha}$}
\caption{ Due to the diurnal motion of the Earth the angle $\Theta$, between the line of observation and the sun's direction of motion, is changing  as the earth rotates. We show the angle $\Theta$ as a function of $\tilde{\alpha}$, essentially time, for various inclinations $\tilde{\delta}$.  The intermediate thickness, the short dash, the long dash, the fine line, the long-short dash, the short-long-short dash and the thick line correspond to inclination $\tilde{\delta}=-\pi/2,-3\pi/10,-\pi/10,0,\pi/10,3\pi/10$ and $\pi/2$  respectively. We see that, for negative inclinations, the angle $\Theta$ can take values near $\pi$, i.e. opposite to the direction of the sun's velocity, where the rate attains its maximum.}
 \label{fig:theta}
\end{center}
  \end{figure}
\section{Diurnal Variation}
Up to now we have considered the event rate in a directional experiment in fixed direction with respect to the galaxy in the galactic system discussed above. The apparatus, of course, will be oriented in a direction specified in the local frame, e.g. by a point in the sky specified, in the equatorial system, by right ascension $\tilde{\alpha}$ and inclination\footnote{We have chosen to adopt the notation $\tilde{\alpha}$ and $\tilde{\delta}$ instead of the standard notation $\alpha$ and $\delta$ employed by the astronomers to avoid possible confusion stemming from the fact that $\alpha$  has already been used to designate the phase of the Earth and $\delta$ has been used for the ratio of the rotational velocity of the Earth around the Sun  by the velocity of the sun around the center of the galaxy} $\tilde{\delta}$. This will lead to a diurnal variation of the event rate \cite{CYGNUS09,Vergados-010}.

The galactic frame, in the so called J2000 system, is defined by the galactic pole with ascension $\tilde{\alpha}_1=12^h~ 51^m~ 26.282^s$ and inclination $\tilde{\delta}_1= +27^0 ~7^{'} ~42.01^{''}$ and the galactic center at $\tilde{\alpha}_2 =17^h ~45^m~ 37.224^s$ , $\tilde{\delta}_2=-(28^0~ 56^{'}~ 10.23^{''} ) $. Thus the galactic unit vector $\hat{y}$, specified by $(\tilde{\alpha}_1,\tilde{\delta}_1)$, and the unit vector $\hat{s}$, specified by $(\tilde{\alpha}_2,\tilde{\delta}_2)$,
can be expressed in terms of the celestial unit vectors $\hat{i}$ (beginning of measuring the right ascension),  $\hat{k}$ (the axis of the Earth's rotation) and $\hat{j}=\hat{k}\times\hat{i}$.
One finds
 \begin{eqnarray}
{\hat y}&=&-0.868{\hat i}- 0.198{\hat j}+0.456{\hat k}
\nonumber \mbox{ (galactic axis) },\\
{\hat x}&=&-{\hat s}=0.055{\hat i}+ 0.873{\hat j}+ 0.483{\hat k} \mbox{ (radially out towards the sun)  },
\nonumber\\
{\hat z}&=&{\hat x}\times{\hat y}=0.494{\hat i}- 0.445{\hat j} +0.747{\hat k}\mbox{ (the sun's direction of motion)}.
\end{eqnarray}
Note in our system the x-axis is opposite to the s-axis used by the astronomers.
Thus a vector oriented by $(\tilde{\alpha},\tilde{\delta}) $ in the laboratory  is given   in the galactic frame by a unit vector with components:
\begin{equation}
\left (
\begin{array}{l}
 y \\
 x \\
 z
\end{array}
\right )
=\left (\begin{array}{l}
 -0.868 \cos {\tilde{\alpha} } \cos {\tilde{\delta} }-0.198 \sin {\tilde{\alpha}
   } \cos {\tilde{\delta }}+0.456 \sin{\tilde{\delta} } \\
 0.055 \cos {\tilde{\alpha}} \cos {\tilde{\delta}}+0.873 \sin {\tilde{\alpha}
   } \cos {\tilde{\delta }}+0.4831 \sin {\tilde{\delta }} \\
 0.494 \cos {\tilde{\alpha} } \cos {\tilde{\delta} }-0.445 \sin {\tilde{\alpha}
   } \cos {\tilde{\delta}}+0.747 \sin {\tilde{\delta} }
\end{array}
\right ),
\end{equation}
where $\tilde{\alpha}_0=282.25^0$ is the right ascension of the equinox, $\gamma\approx 33^0$ was given above and $\theta_P=62.6^0$ is the angle the Earth's north pole forms with the axis of the galaxy. \\Due to the Earth's rotation the unit vector $(x,y,z)$, with a suitable choice of the initial time, $\tilde{\alpha}-\tilde{\alpha}_0=2\pi(t/T)$, is changing as a function of time

 The  angle $\Theta$  scanned by the direction of observation is shown, for various inclinations $\tilde{\delta}$, in Fig.~3. We see that for  negative inclinations, the angle $\Theta$ can take values near $\pi$, i.e. opposite to the direction of the sun's velocity, where the rate attains its maximum.

\begin{figure}
\begin{center}
 {
\rotatebox{90}{\hspace{0.6cm} {$\kappa $ (sense known)}}
\includegraphics[scale=0.5]{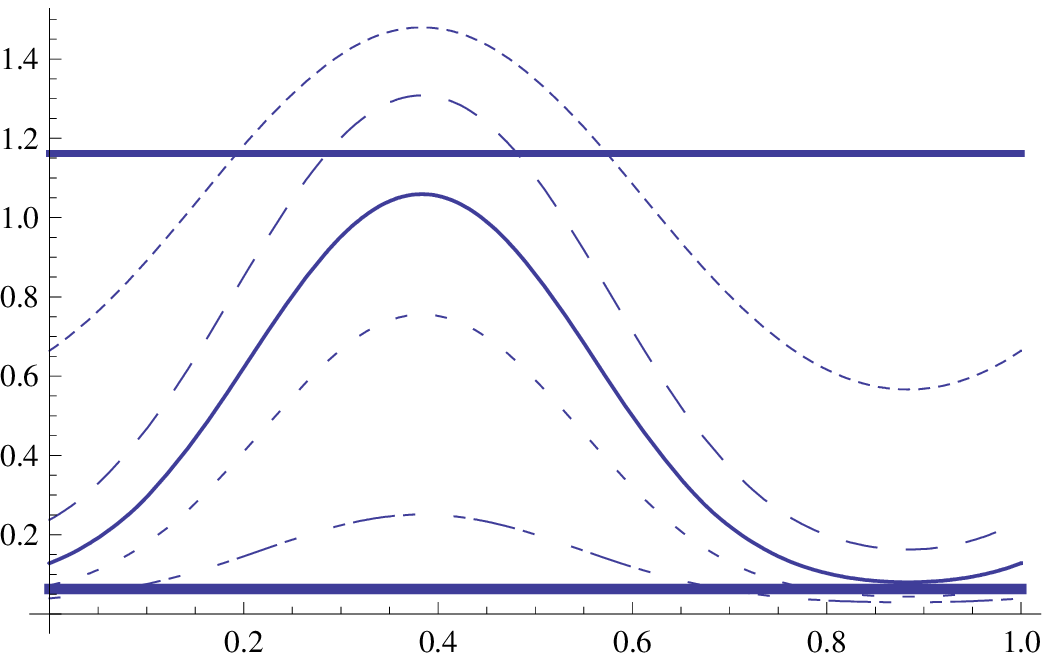}
}
 \hspace{1.0cm}
 {
\rotatebox{90}{\hspace{0.6cm} { $\kappa $ (both sense)     }  }
\includegraphics[scale=0.5]{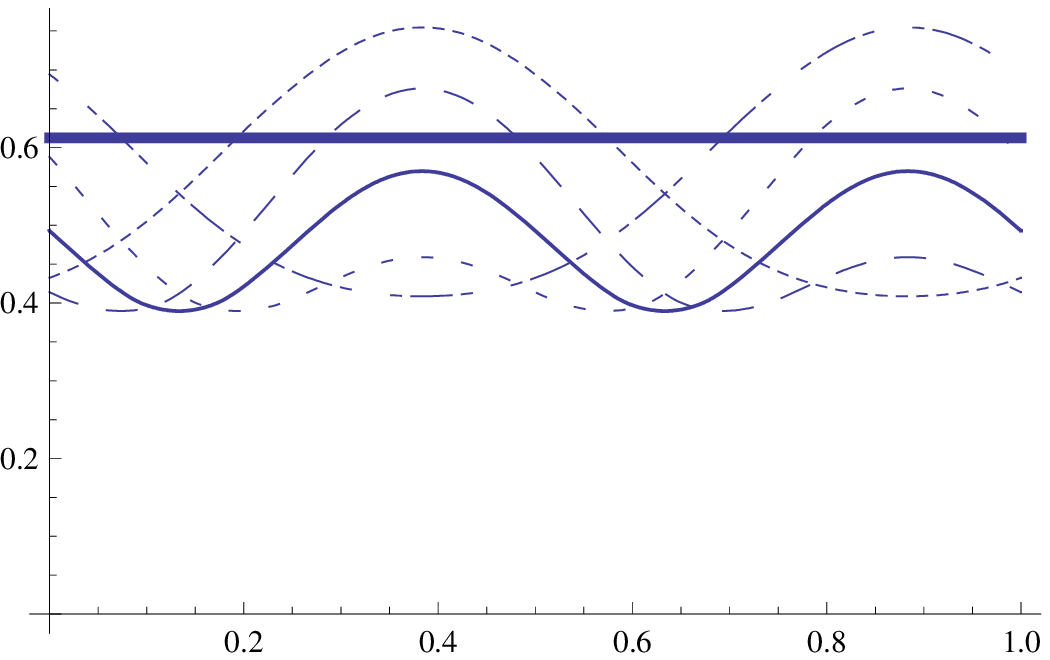}
}
\\
\hspace{-0.0cm} {$\frac{t}{T} \longrightarrow$}
\caption{The time dependence(in units of the Earth's rotation period)   of the parameter $\kappa$  for various inclinations $\tilde{\delta}$ in the case of the target CS$_2$. The intermediate thickness, the short dash, the long dash, the fine line, the long-short dash, the short-long-short dash and the thick line correspond to inclination $\tilde{\delta}=-\pi/2,-3\pi/10,-\pi/10,0,\pi/10,3\pi/10$ and $\pi/2$  respectively.   Due to the diurnal motion of the Earth different angles $\Theta$ are sampled as the earth rotates.
}
 \label{fig:Sdiurnal}
 \end{center}
  \end{figure}
\begin{figure}
 \begin{center}
 {
\rotatebox{90}{\hspace{0.6cm} { $\kappa $ (sense known)}}
\includegraphics[scale=0.5]{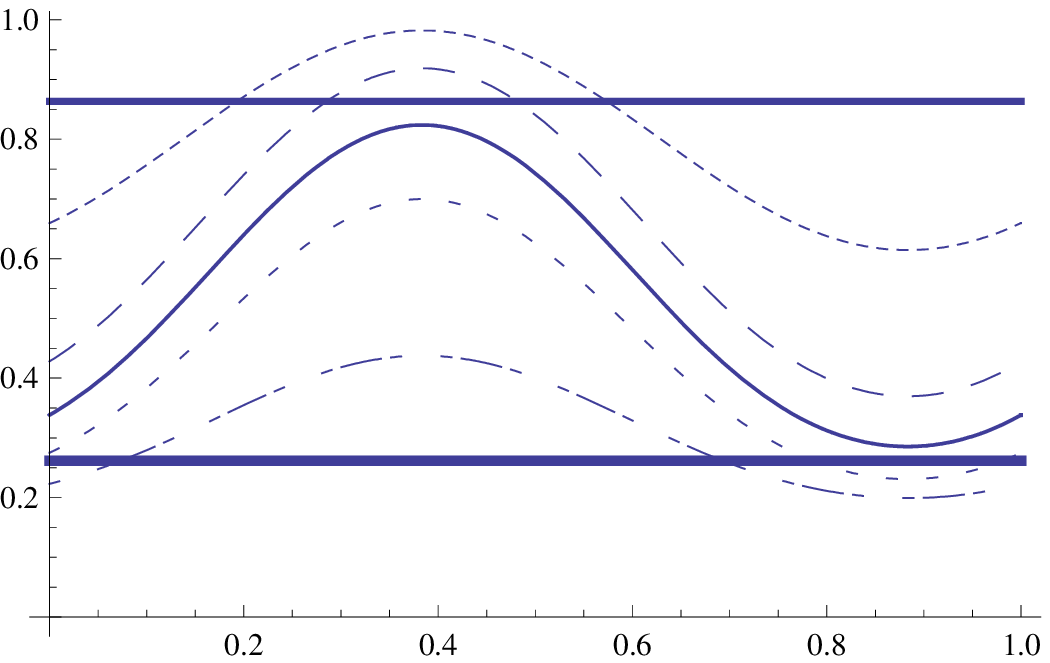}
}
 \hspace{1.0cm}
 {
\rotatebox{90}{\hspace{0.6cm} { $\kappa $ (both senses)}}
\includegraphics[scale=0.5]{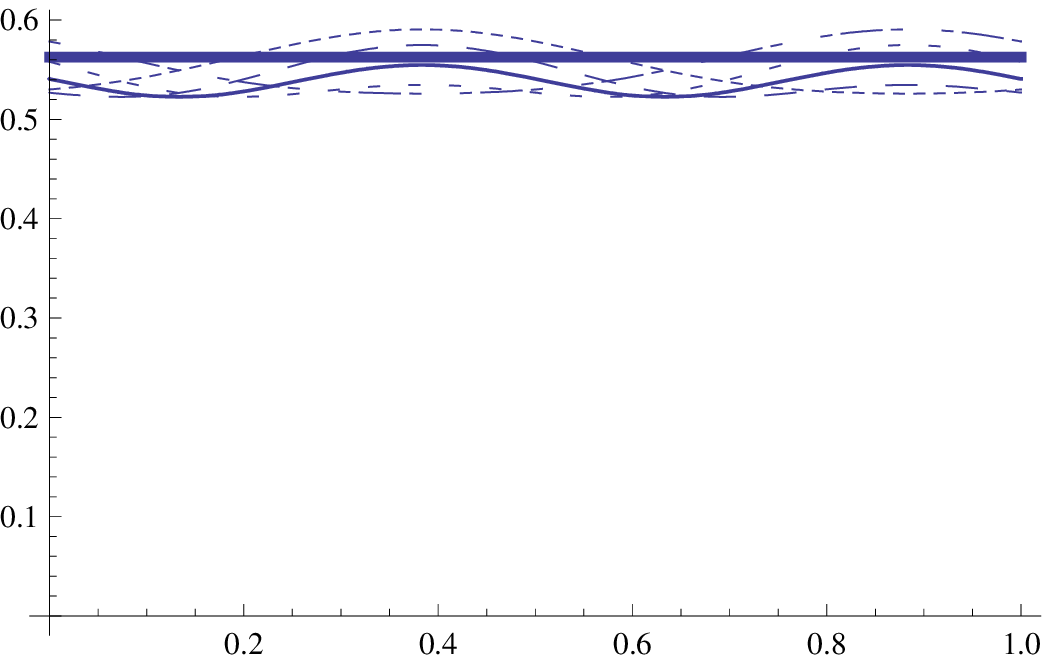}
}
\\
\hspace{-0.0cm} {$ \frac{t}{T} \longrightarrow$}
\caption{The same  as in Fig. \ref{fig:Sdiurnal} in the case of the Iodine target.
The intermediate thickness, the short dash, the long dash, the fine line, the long-short dash, the short-long-short dash and the thick line correspond to inclination $\tilde{\delta}=-\pi/2,-3\pi/10,-\pi/10,0,\pi/10,3\pi/10$ and $\pi/2$  respectively. }
 \label{fig:Idiurnal}
  \end{center}
  \end{figure}

The equipment scans different parts of the galactic sky, i.e. observes different angles $\Theta$. So the rate will change with time depending on whether the sense of the recoiling nucleus can be determined along the line of recoil. The results depend, of course, on the WIMP mass and the target employed. We will consider a light (the time dependence of $\kappa$  is exhibited in Fig. \ref{fig:Sdiurnal}) and an intermediate-heavy target (the time dependence of $\kappa$  is exhibited in Fig. \ref{fig:Idiurnal}).


 \section{Discussion}
In directional experiments one measures not only the  energy of the recoiling nucleus but also
the direction the nucleus is recoiling.
Some of the requirements that should be met by such detectors have recently been discussed \cite{CKS-DS05,Green06}.
To fully exploit the advantages of such detectors one should be able to distinguish between recoils with
momenta ${\bf p}$ and $-{\bf p}$ (sense of direction), which now appears to be feasible.
Some of the predicted interesting features of directional event rates persist, however, even if it turns out
that the sense of motion of recoils along their line of motion cannot be measured.
Such experiments, given a sufficient number of events, provide an excellent signature to discriminate against background.
One expects, of course,  a smaller rate by observing  in a given direction. In the most favored direction, opposite to the sun's direction of
motion, the event rate is $\approx \frac{\kappa}{2 \pi}$ down from that of the
standard non directional experiments, if a specific angle $\Phi$ is chosen.  Since, however, $\kappa$ is independent of the angle $\Phi$, one can integrate over all azimuthal angles and thus the retardation is only of order $\kappa$ ($\kappa\approx1$ in the most favored direction).The parameter $\kappa$ is essentially independent of any particle model parameters other than the WIMP mass. It does depend, however, on the assumed velocity distribution and to the nuclear form factor.
Finally we have seen that in directional experiments the relative event rate for detecting WIMPs within our galaxy, as given by the parameter $\kappa$, will show a periodic diurnal variation due to the rotation of the Earth.  The time variation is larger in the case of light WIMP and/or light target. So from this perspective the lighter target is preferred.
The time variation arising after the inclusion of the modulation parameters ($h_m, \alpha$) or ($h_c,h_s$) is expected to be even more complicated, since these parameters depend on both $\Theta$ and $\Phi$. One expects a diurnal variation on top of the annual variation characteristic of the usual modulation effect entering both directional and non directional experiments. Such effects will be discussed elsewhere.

\subsection*{Acknowledgments}
The work of one of the authors (JDV) was partially supported by the  grant MRTN-CT-2006-035863 (UniverseNet). He is indebted to Professors N. Spooner and J. Jochum for useful discussions
on experimental aspects of this work and David Delepine for his hospitality in Leon during DSU10.


\end{document}